\documentclass[final,3p,times]{elsarticle}

 \usepackage{graphicx}

\usepackage{amssymb,bm,mathrsfs,bbm,amscd}
\usepackage{booktabs}
\biboptions{numbers,sort&compress}
\usepackage[shortlabels]{enumitem}
\setlist[enumerate]{nosep}
\usepackage{float}
\usepackage{color}

\usepackage{lineno}

\begin{document}

\begin{frontmatter}



\title{Strongly Resonating Bosons in Hot Nuclei}


\author[label1]{S. Zhang}
\ead{zsylt@imun.edu.cn}
\author[label2,label3]{A. Bonasera}
\author[label1]{M. Huang}
\ead{huangmeirong@imun.edu.cn}
\author[label3,label4]{H. Zheng}
\author[label5]{G. Zhang}
\author[label2,label6]{Z. Kohley}
\author[label1]{L. Lu}
\author[label5]{Y.G. Ma}
\author[label2,label6]{S.J. Yennello}
\address[label1]{College of Physics and Electronics information, Inner Mongolia University for Nationalities, Tongliao, 028000, China.}
\address[label2]{Cyclotron Institute, Texas A$\&$M University, College Station, Texas 77843, USA.}
\address[label3]{Laboratori Nazionali del Sud, INFN, via Santa Sofia, 62, 95123 Catania, Italy.}
\address[label4]{School of Physics and Information Technology, Shaanxi Normal University, Xi'an 710119, China.}
\address[label5]{Shanghai Institute Applied Physics, Chinese Academy of Sciences, Shanghai 201800, China.}
\address[label6]{Chemistry Department, Texas A$\&$M University, College Station, Texas 77843, USA.}

\begin{abstract}
When two heavy ions near the Fermi energy collide, a warm and low-density region can form in which fragments appear. This region is mainly dominated by proton ($p$) and alpha ($\alpha$) particles. In such an environment, the $\alpha$s interact with each other, and especially through strong resonances, form complex systems such as $^{8}$Be and $^{12}$C. Our experiments show that in the reactions $^{70(64)}$Zn($^{64}$Ni)+$^{70(64)}$Zn($^{64}$Ni) at E/A=35 MeV/nucleon levels of $^{8}$Be appear around relative energies E$_{ij}$=0.092 MeV, 3.03 MeV as well as above 10 MeV and 100 MeV. For the 3$\alpha$ systems, multi resonance processes give rise to excited levels of $^{12}$C. In particular, the Hoyle state at 7.654 MeV excitation energy shows a decay component through the ground state of $^{8}$Be and also shows components where two different $\alpha$ couples are at relative energies consistent with the ground state of $^{8}$Be at the same time. A component where the three $\alpha$ relative energies are consistent with the ground state of $^{8}$Be (i.e., E$_{12}$=E$_{13}$=E$_{23}$=0.092 MeV) is also observed at the 7.458 MeV excitation energy, which was suggested as an Efimov state.
\end{abstract}

\begin{keyword}
Hoyle state \sep Efimov state \sep Sequential decay \sep Direct decay \sep Heavy ion reactions
\end{keyword}

\end{frontmatter}


Several decades after the suggestion by Fred Hoyle \cite{Hoyle:1954} of a 0$^{+}$ resonance near the 3$\alpha$ threshold to accelerate $^{12}$C formation in stars, the Hoyle state (HS) is still a hot topic in nuclear structure
\cite{Marin-Lambarri:2014zxa,Epelbaum:2011md,Enyo:2007zz,Ishikawa:2014mza,Morinaga:1966pl,Freer2014PPNP,
Freer:1994zz,Raduta:2011yz,Kirsebom:2012zza,Manfredi:2012zz,Rana:2013hka,Itoh:2014mwa,DellAquila:2017ppe,Smith:2017jub,Kirsebom:view2017}.
While its energy, i.e., 7.654 MeV, and width, 8.5 eV, are firmly established, there are debates on its decay. It is commonly accepted \cite{Ishikawa:2014mza,Morinaga:1966pl,Freer2014PPNP,Freer:1994zz,Raduta:2011yz,Kirsebom:2012zza,Manfredi:2012zz,Rana:2013hka,
Itoh:2014mwa,DellAquila:2017ppe,Smith:2017jub,Kirsebom:view2017} that almost 100\% of the HS decay is through the ground state of $^{8}$Be ($^{8}$Be$_{g.s.}$), thus corresponding to a sequential decay (SD), i.e., first an $\alpha$ particle is emitted, and then $^{8}$Be decays into 2$\alpha$s with 0.092 MeV relative kinetic energy. Other decay modes, for example, the theoretically predicted direct decay (DD) of $^{12}$C into 3$\alpha$s of equal energy (DDE) or into a linear chain (LD) \cite{Enyo:2007zz,Ishikawa:2014mza} have been studied in high precision/high statistical experiments \cite{Raduta:2011yz,Rana:2013hka,Manfredi:2012zz,DellAquila:2017ppe,Smith:2017jub,Kirsebom:view2017,Kirsebom:2012zza,Itoh:2014mwa,Freer:1994zz}
giving an upper limit of 0.043\%\cite{Smith:2017jub}, 0.036\%\cite{DellAquila:2017ppe} and 0.024\%\cite{DellAquila:2017ppe} for DD, DDE, and LD, respectively. While we are probably at the limit of the experimental sensitivity, higher statistical experiments might be performed, or different strategies might be explored. In this paper, we will discuss a completely new approach, i.e., we will generally explore the $^{12}$C decays also in the presence of nearby nuclear matter. This is surely relevant since stellar processes, where $^{12}$C (or larger nuclei) are formed, might occur inside a dense star or on its surface, thus occurring under different conditions of density and temperature. One way to simulate some stellar conditions is to collide two heavy ions at beam energies near the Fermi energy.  In central/peripheral collisions of the two ions, first we have a gentle increase in the density slightly above the ground state density, $\rho_{0}$=0.16$fm^{-3}$ \cite{PPNP,Bonasera:1994pr} as revealed by microscopic calculations and experiments \cite{Suno:2016fjb, Schmidt:2016lpt, Mabiala:2016gpt, Marini:2015zwa}. The system expands while cooling, and for densities below (1/3-1/6)$\rho_{0}$, clusters start to appear; this is referred to as the freeze-out region. Not only are protons and alpha particles formed, but also nuclei of larger masses with neutron (N) and proton (Z) contents. Under such conditions, excited fragments might decay, and nearby small fragments might coalesce and form new nuclei. The study of the formation and decay of complex fragments in nuclear matter is of interest, and in particular, in this work, we will study the decay modes of $^{12}$C and $^{8}$Be in 3 and 2$\alpha$s, respectively. To increase the statistics, we combined the results of 3 different experiments: $^{70}$Zn+$^{70}$Zn, $^{64}$Zn+$^{64}$Zn, and $^{64}$Ni+$^{64}$Ni all at 35 MeV/nucleon \cite{Kholey:2010phd} and checked that each system separately produced results in agreement with the others at least for the observables discussed in this paper.
There are many reasons for studying decays in nuclear fragmentation, in particular:
\begin{enumerate}
\renewcommand{\labelenumi}{\theenumi)}
\setlength{\itemsep}{0pt}
\setlength{\parsep}{0pt}
\setlength{\parskip}{0pt}
\item In nuclear matter, the levels and widths of decaying nuclei might be shifted/modified because of the interaction with nearby species. For example, a change in the width of a resonance as compared to the vacuum might tell us the time duration of the freeze-out region.
\item Levels not observed in a vacuum might appear in the surrounding medium, for example, if many $\alpha$s are formed at relative kinetic energies where E$_{ij}$=0.092 MeV, then strong resonances among these bosons could give rise to correlations and Bose Einstein Condensation (BEC) \cite{Efimov:1970zz,Marini:2015zwa}. These conditions might be identified with the strong resonance region in microscopic calculations of the structure of $^{12}$C \cite{Ishikawa:2014mza,Schmidt:2016lpt}. If this is a new state of $^{12}$C as yet unobserved, it could be assigned as an Efimov state (ES). This is a general feature first predicted for nuclear systems
    \cite{Efimov:1970zz, efimov:nature09} and only observed in atomic systems \cite{Greene:2010PT, Zaccanti:2009NP} before our work was conducted.
\item Are the decay modes of a particular resonance modified in medium? For example, is the SD of the HS still dominant with respect to the DD?
\item The SD dominance of the HS tells us that the relative kinetic energy of two $\alpha$s is 0.092 MeV. What is the relative kinetic energy of the third  $\alpha$ with respect to the first two? Why is the HS located at 7.654 MeV?
\end{enumerate}

These and many other questions prompted us to follow an unconventional approach to nuclear structure and decay. The important tool is the detector, which must measure at a high precision the energy and angle of fragments ejected from heavy-ion collisions. We analyzed the data from an experiment performed at the Cyclotron Institute, Texas A$\&$M University \cite{Kholey:2010phd, Mabiala:2016gpt}, using the NIMROD 4$\pi$ detector \cite{NIMROD}. Experimental details can be found in the references \cite{Schmidt:2016lpt, Mabiala:2016gpt, Kholey:2010phd, NIMROD}. Here, it suffices to discuss the approach we follow in this paper. The detector measures the charge Z and mass A of each fragment up to Z=30 and A=50 on an event-by-event basis, see supplemental material for more details. From all the collected events, we selected 3$\alpha$ events, and the analysis was performed only on them. The momenta of the $\alpha$s are rather well measured with the only major problem arising when the relative kinetic energy of the 2$\alpha$s is small, of the order of tens keV. Such particles are detected in two nearby detectors (or in the same one) of NIMROD, and because of the detector's finite granularity, an error results in the angle and consequently for the relative momenta. This is a problem for all detector types and the resulting error or minimum relative kinetic energy which can be measured is of the order of 40 keV \cite{Raduta:2011yz, Freer:1994zz}; the larger the granularity, the smaller the error. An evident advantage of our method, since we are working at the beam energy close to the Fermi energy, is that each ion has high kinetic energy, larger than few MeV/nucleon, see Fig. S3 in supplement. This is optimal for our detector. It is the relative kinetic energy between $\alpha$ particles that needs to be small in order to reveal low energy excited levels of $^{8}$Be and $^{12}$C.

We emphasize that only the events with $\alpha$ multiplicity equal to three are analyzed in our analysis. Let us start by recalling the relation that calculate the excitation energy E${^*}$ of $^{12}$C decaying into 3$\alpha$s with Q-value, Q=-7.275 MeV.
This relationship is given by Eq. (1):
\begin{small}
\begin{equation}
E^* = \frac{2}{3}\sum^3_{i=1,j>i} E_{ij} - Q,
\label{Estar}
\end{equation}
\end{small}
where E$_{ij}$ is the relative kinetic energy of 2$\alpha$ and is measured event-by-event. Thus, we can easily estimate Eq.(\ref{Estar}).
Notice that the important ingredients entering Eq.(\ref{Estar}) are the relative kinetic energies; since we have three indistinguishable bosons, we analyze the E$_{ij}$ distribution by cataloging for each event the smallest relative kinetic energy, E$^{Min}_{ij}$, the middle relative  kinetic energy, E$^{Mid}_{ij}$, and the largest relative kinetic energy, E$^{Lar}_{ij}$.

\begin{figure}[h]
\vspace{-5pt}
\setlength{\abovecaptionskip}{0.cm}
\setlength{\belowcaptionskip}{-0.cm}
\centering
\includegraphics[scale=0.65]{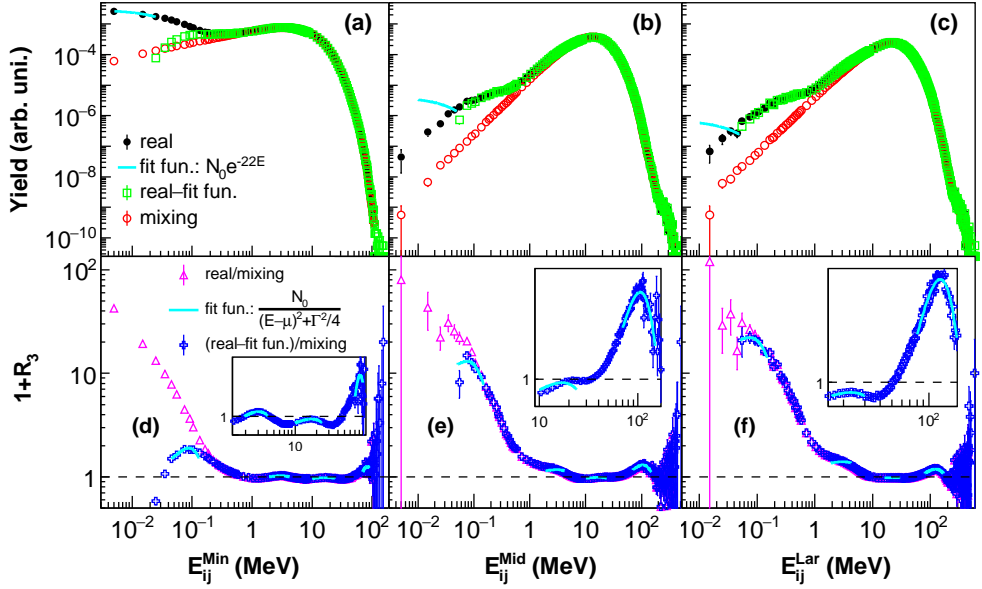}
\caption{\footnotesize (color online) Selected events from $^{70(64)}$Zn($^{64}$Ni)+$^{70(64)}$Zn($^{64}$Ni) at E/A=35 MeV/nucleon with $\alpha$ multiplicity equal to three. (a) Relative kinetic energy distribution as a function of the minimum relative kinetic energy, (b) the middle relative kinetic energy and (c) the largest relative kinetic energy of 2$\alpha$s. The solid black circles represent data from real events, red open circles are from mixing events, and the green open squares represent the difference between the real events and the exponential function (solid line), which takes into account the experimental error.
The ratios of the real (pink open triangles) data and the real data minus the fitting function (blue crosses) divided by the mixing events are, respectively, a function of (d) the minimum relative kinetic energy, (e) the middle relative kinetic energy and (f) the largest relative kinetic energy of 2$\alpha$s. The solid lines are Breit-Wigner fits.}
\label{RelE}
\vspace{-15pt}
\end{figure}

In Fig. \ref{RelE}, we plot the relative kinetic energy distributions for these three cases.
In the top panels, the solid black circles give the distributions obtained from the real events. They show bumps but no real structures.
This is due to the fact that in the fragmentation region, some $\alpha$ may come from the decay of $^{8}$Be or $^{12}$C or
they might come from completely non-correlated processes, for example, the $\alpha$ emission from a heavy fragment.
To distinguish the correlated from the non-correlated events, we randomly choose three different $\alpha$ from three different events
and build the distributions displayed in Fig. \ref{RelE} (mixing events-red open circles).
The total number of real and mixing events are normalized to one, respectively.
The two distributions look similar in log-scale but show some remarkable differences at low relative kinetic energies. While the distribution of the real events in Fig. \ref{RelE}(a) goes down, it goes up in Fig. \ref{RelE}(b) and (c). As we have mentioned, when the relative kinetic energy becomes very small, it becomes difficult to assign the angle of detection because of the detector granularity. This is, of course, less important for the Fig. \ref{RelE}(b) and (c) cases since the smallest relative kinetic energies are obtained for the events in the first panel. To correct this feature, we fit the highest points of Fig. \ref{RelE}(a) with an exponential function. This allows us to derive the instrumental error $\Delta$E=1/22 MeV=0.045 MeV, see Fig. \ref{RelE}(a). The estimated error is slightly larger than what is found in the refs. \cite{Raduta:2011yz,Rana:2013hka,Manfredi:2012zz,DellAquila:2017ppe,Smith:2017jub,Kirsebom:view2017,Kirsebom:2012zza,Itoh:2014mwa,Freer:1994zz}
but small enough to let us derive the results we discuss below. For Fig. \ref{RelE}(b) and (c), it is evident that the experimental error is less important and a change in slope can be seen around 0.1 MeV (the $^{8}$Be$_{g.s.}$). To uncover the resonance, we can perform an exponential fit using the same slope (or experimental error) as before and fit the experimental point at 0.045 MeV. By subtracting the fits from the real events, we obtain the open squares in Fig. \ref{RelE}, which can be considered as the real events corrected by the detector acceptance. As we can see, all three cases display a bump around 0.08 MeV (very close to 0.092 MeV) corresponding to the decay of $^{8}$Be$_{g.s.}$. Because of the way we have ordered the 2$\alpha$s relative kinetic energies, we can deduce that if the largest relative kinetic energy is 0.092 MeV, the other two must be 0.092 MeV as well (since they are smaller). Thus, we have events where the 3$\alpha$s are in mutual resonance, which is a mechanism similar to the Efimov states where a Boson is exchanged between the other two \cite{Efimov:1970zz,Naidon:2017dpf,efimov:nature09}. If this cannot be associated with a feature of a strongly resonating Boson gas \cite{Ishikawa:2014mza}, it might be an unexplored state of $^{12}$C at E$^*$=7.458 MeV (as given by Eq.(\ref{Estar})). In ref. \cite{Zheng:2018plb}, the possibility of such a state was discussed, and it was suggested that its observation probability is 8-orders of magnitude smaller than the HS. Of course those calculations were in a vacuum; when in a medium, the mechanism of mutual resonances might be enhanced in the presence of other fragments, which could be more complex Bosons (for example, $^{16}$O) or Fermions (for example, $^{7}$Li). Taking the ratio of the real (minus the exponential fit) and the mixing events gives the correlation functions 1+R$_3$ displayed in the bottom panels of Fig. \ref{RelE}. The ratio shows that not only are peaks present around 0.08 MeV for all cases, but also at 3.05 MeV corresponding to the 1st excited level of $^{8}$Be for the smallest relative kinetic energies, and higher energy peaks are also visible.

\begin{table}[!b]
\vspace{-12pt}
\setlength{\abovecaptionskip}{0.cm}
\setlength{\belowcaptionskip}{-0.cm}
   \centering
   \scriptsize
   \tabcolsep=3.5pt
   \caption{\footnotesize The parameters of BW fit to peaks in Fig. \ref{RelE}}
   \vspace{-6pt}
   \begin{tabular}{cccc}\\
   \hline \hline
             &E$^{Min}_{ij}$: $\mu$(MeV), $\hbar/\Gamma$(fm/c)& E$^{Mid}_{ij}$: $\mu$(MeV), $\hbar/\Gamma$(fm/c)& E$^{Lar}_{ij}$: $\mu$(MeV), $\hbar/\Gamma$(fm/c)\\
   \hline
             peak${1}$ & 0.088$\pm$0.001, 1192$\pm$66  & 0.08$\pm$0.02, 1089$\pm$288  & 0.08$\pm$0.04, 984$\pm$540   \\
             peak${2}$ & 3.05$\pm$0.01, 14.2$\pm$0.3   &                              &                              \\
             peak${3}$ & 17.0$\pm$0.1, 2.08$\pm$0.04   &                              & 22.9$\pm$0.3, 1.1$\pm$0.1    \\
             peak${4}$ & 83$\pm$3, 2.8$\pm$1.0         & 106$\pm$1, 0.95$\pm$0.04     & 124.1$\pm$0.9, 0.70$\pm$0.02  \\

   \hline \hline
   \end{tabular}
   \label{tab1}
   \vspace{-6pt}
   \end{table}

A Breit-Wigner (BW) fit to these resonances gives the parameters reported in Table ~\ref{tab1}. No clear peaks can be seen for the 3.05, and 17.0 MeV cases in Fig. \ref{RelE}(e) and (f). We used the same fitting values obtained in Fig. \ref{RelE}(d) for the 3.05, and 17.0 MeV cases in
Fig. \ref{RelE}(e) and for the 3.05 MeV case in Fig. \ref{RelE}(f). For the largest relative kinetic energy case, Fig. \ref{RelE} (f), a peak appears around 23 MeV. As we can see, the centroids of the resonances are in good agreement with values in a vacuum, but the widths are somehow larger either due to the experimental acceptance or, most probably, to the presence of a medium. These differences from the vacuum values are more marked for higher energies in the region 10-20 MeV where we know there are many resonances in $^{8}$Be. Some of these resonances have large widths and others have narrow widths \cite{NNDC}. Probably, when in a medium, the effects in the freeze-out region are responsible for this feature, and if the width is associated with the lifetime of the system, then we can derive the value $\tau=\hbar/\Gamma\cong1$ fm/c. There are no reported levels of $^{8}$Be (or $^{12}$C) around 100 MeV; thus, we could associate these bumps with some or many unobserved new levels at high excitation energies. However, it should be noted that the corresponding excitation energy per particle of $^{12}$C: E$^*$=100/12 MeV =8.33 MeV is very close to the center of mass energy in the heavy-ion collisions E$_{cm}$=35/4 MeV=8.75 MeV. Thus the correlation at such high energies might be due to the formation of a hot Boson-Fermion gas mixture and has been analyzed in some detail in refs. \cite{Schmidt:2016lpt, Mabiala:2016gpt, Marini:2015zwa}. The lifetime of such a system is of the order of 1 fm/c. It is interesting to notice that the quantity 1+R$_3$ is smaller than one in the energy region above 10 MeV and becomes larger than one around 100 MeV (recall that the real and mixing events are normalized to the same area).

\begin{figure}[h]
\vspace{-5pt}
\setlength{\abovecaptionskip}{0.cm}
\setlength{\belowcaptionskip}{-0.cm}
\centering
\includegraphics[scale=0.6]{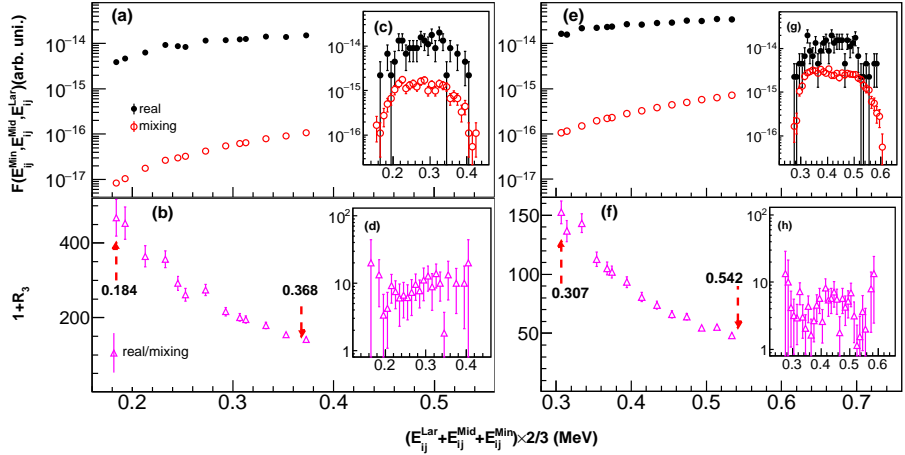}
\caption{\footnotesize (color online) The distribution of the F(E$^{Min}_{ij}$, E$^{Mid}_{ij}$, E$^{Lar}_{ij}$) and 1+R$_3$ as a function of the total relative kinetic energy. The solid black circles are from the real events, red open circles are the mixing events, pink open triangles indicate the ratios of the real and the mixing events. The red dashed arrows show the minimum and maximum values of the correlation function allowed by the triangular inequality.
Left panels (a, b) correspond to E$^{Min}_{ij}$= E$^{Mid}_{ij}$=0.092 MeV, right panels (e, f) correspond to E$^{Min}_{ij}$=0.092 MeV, E$^{Mid}_{ij}$=0.092$\times$2 MeV. In the insets, the total relative kinetic energy distributions for the real (mixing) events and their ratio with the selections E$^{Min}_{ij}$=E$^{Mid}_{ij}$=0.092$\pm$0.06/3 MeV (c, d), and the selections E$^{Min}_{ij}$=0.092$\pm$0.06/3, E$^{Mid}_{ij}$=0.092$\times$2$\pm$0.06/3 MeV (g, h).}
\label{yield_2}
\vspace{-5pt}
\end{figure}

To understand the role of $^{8}$Be in the decay of $^{12}$C in a medium, we define the product of the yield distributions displayed in Fig. \ref{RelE} as Eq.(\ref{yield}):
\begin{small}
\begin{equation}
 F(E^{Min}_{ij}, E^{Mid}_{ij}, E^{Lar}_{ij}) = Y(E^{Min}_{ij})Y(E^{Mid}_{ij})Y(E^{Lar}_{ij}).
 \label{yield}
\end{equation}
\end{small}
Theoretically, since this quantity can be obtained from the real correlation functions plotted in Fig. \ref{RelE}, it is the exact (apart from the experimental error) three-body correlation function or the average three-body correlation function if it is built from the mixing events \cite{Lifschitz}. Of course there are an infinite number of possible choices for the relative kinetic energies, but we will restrict ourselves to a few illustrative and instructive cases. In Fig. \ref{yield_2}(a), we plot the quantity F(0.092 MeV, 0.092 MeV, E$^{Lar}_{ij}$), i.e., the two smallest relative kinetic energies are consistent with the $^{8}$Be$_{g.s.}$ energy. When two relative energies are fixed, the third one must fulfill the triangular inequality. Thus, the value of total relative kinetic energy should be constrained between 0.184 MeV $\le$ (E$^{Lar}_{ij}$+E$^{Mid}_{ij}$+ E$^{Min}_{ij}$)$\times$$\frac{2}{3}$ $\le$ 0.368 MeV, arrows in the Fig. \ref{yield_2}(a). The results from the real (solid black circles) and the mixing events (red open circles) are given as function of the kinetic contribution to the $^{12}$C excitation energy given in Eq.(\ref{Estar}). There is a large yield difference between the real and the mixing events and their ratio depicted in Fig. \ref{yield_2}(b) is much larger than one. The 1+R$_3$ quantity has a broad bump around 0.18 MeV corresponding to three equal energies, i.e., the ES or a resonating Boson gas in a nuclear medium. A change of slope around 0.38 MeV corresponding to the HS is visible; in this case, the largest relative kinetic energy of the two $\alpha$ particles is about 0.368 MeV, i.e., four times the $^{8}$Be$_{g.s.}$ energy. In the right panels of Fig. \ref{yield_2}(e) and (f), we plot the results corresponding to F(0.092 MeV, 0.092$\times$2 MeV, E$^{Lar}_{ij}$), i.e., the middle relative kinetic energy is twice of the $^{8}$Be$_{g.s.}$ energy again under the triangular inequality restriction, arrows in the Fig. \ref{yield_2}(b). These cases technically correspond to SD, it is possible that two of the three $\alpha$ relative kinetic energies are consistent with the $^{8}$Be$_{g.s.}$ or with multiples of the $^{8}$Be$_{g.s.}$ energy. This might call to remind ternary classical systems that become unstable when their frequencies are multiples of a fundamental frequency, for example, the occurrence of asteroids in the Sun-Jupiter belt at frequency multiples of the Jupiter procession around the Sun \cite{fluid}.

In order to strengthen these results, we plotted the total relative kinetic energy distributions of the real (mixing) events and their ratio with the selections E$^{Min}_{ij}$=E$^{Mid}_{ij}$=0.092$\pm$$\delta$E/3 MeV, and the selections E$^{Min}_{ij}$=0.092$\pm$$\delta$E/3 MeV, E$^{Mid}_{ij}$=0.092$\times$2$\pm$$\delta$E/3 MeV. Since the error on the largest relative kinetic energy is smaller than the other relative combinations, see Fig. \ref{RelE}, we can decrease $\delta$E at the statistical limit. In Fig. \ref{yield_2}, we have included the results when $\delta$E=0.06 MeV. In such cases, only 90 (Fig. \ref{yield_2}(c, d)) and 130 (Fig. \ref{yield_2}(g, h)) real events survive, respectively. Other cases with different $\delta$E can be found in the supplemental material, Fig. S4 and S5. The results are in agreement with the correlation function, Eq.(\ref{yield}). A dedicated experiment with higher statistics and better detector system, say for $^{40}$Ca+$^{40}$Ca collisions around 40 MeV/nucleon, should shed further light on the properties of the resonating bosons in hot matter.

\begin{figure}[h]
\vspace{-5pt}
\setlength{\abovecaptionskip}{0.cm}
\setlength{\belowcaptionskip}{-0.cm}
\centering
\includegraphics[scale=0.45]{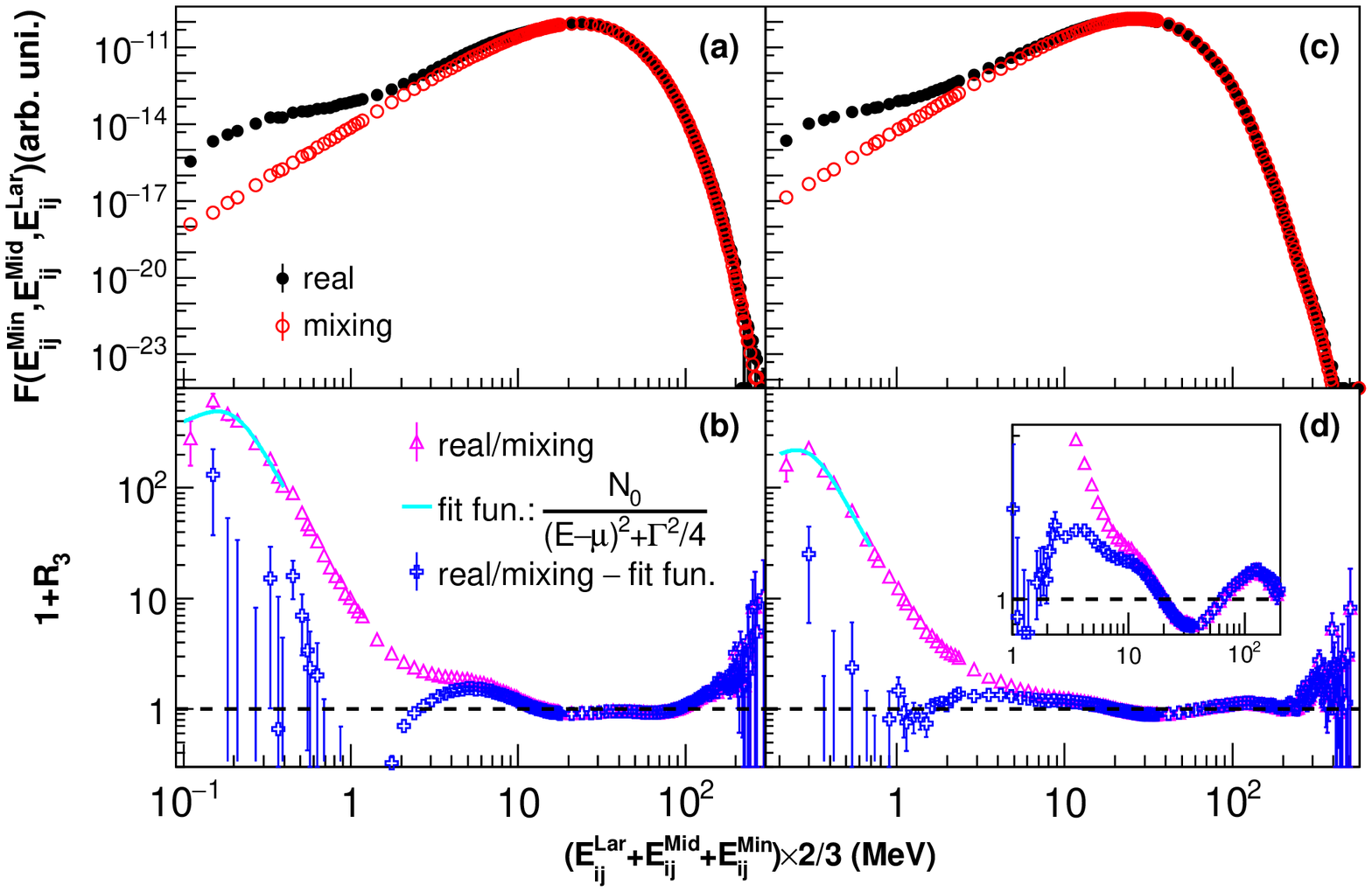}
\caption{\footnotesize (color online) The distribution of the F(E$^{Min}_{ij}$, E$^{Mid}_{ij}$, E$^{Lar}_{ij}$) and 1+R$_3$ as a function of the total relative kinetic energy. The solid black circles are the real events, red open circles are mixing events, pink open triangles indicate the ratios of the real and mixing events, and blue crosses are  difference between triangles and the fit function (solid line). Left panels (a, b) correspond to E$^{Min}_{ij}$= E$^{Mid}_{ij}$=E$^{Lar}_{ij}$, and right panels (c, d) correspond to E$^{Min}_{ij}$=E$^{Mid}_{ij}$=E$^{Lar}_{ij}/4$.}
\label{yield_3}
\vspace{-5pt}
\end{figure}

Other interesting cases are obtained when the 3$\alpha$s relative kinetic energies are equal (DDE), Fig. \ref{yield_3}-left panels, or two energies are equal and the third one is twice the sum of the first two (LD), Fig. \ref{yield_3}-right panels. A peak is observed, as expected, at low relative kinetic energies in the left panels corresponding to the ES. Notice that the condition of the three equal relative kinetic energies naturally produces a resonance at low relative kinetic energies. A BW fit gives a centroid at 0.16$\pm$0.02 MeV (close to the theoretical 0.184 MeV) and with 0.24$\pm$0.02 MeV width,
see Fig. \ref{yield_3}(b). If the width is not an experimental effect then it would correspond to a lifetime of the resonance of about 1000 fm/c which is a typical evolution time of an heavy ion collision. The ratio is less than 1 below 3 MeV, which means, that the 9.641 MeV and the 10.3 MeV have no DDE contributions. These findings are consistent with the conclusions given by Raduta et al. \cite{Raduta:2011yz}. The absence of such decay modes cannot be due to effects of a medium, which we believe will enhance the possibility of three-equal energy decays and should be confirmed with experiments, where the effects of a medium are not present. The LD displayed in Fig. \ref{yield_3}(d) shows a large bump as well which could be well fitted with a BW distribution. The fit gives the centroid 0.25$\pm$0.02 MeV and a width 0.35$\pm$0.02 MeV. These values suggest that this component could be present in the HS as well. In the next paragraph, we will discuss the contribution of this configuration to the HS and compare to other experimental results.

In principle, the observables displayed in Figs. \ref{yield_2} and \ref{yield_3} might be directly obtained from the data, but the experimental error and the poor statistics do not allow us to do so, as shown in the insets of Fig. \ref{yield_2} (c, d, g and h). Thus, we derived another observable, which gives the probability of decay of $^{12}$C into a particular mode (say DDE or LD) with respect to SD. These probabilities have been discussed experimentally using different techniques \cite{Raduta:2011yz,Rana:2013hka,Manfredi:2012zz,DellAquila:2017ppe,Smith:2017jub,Kirsebom:view2017,Kirsebom:2012zza,Itoh:2014mwa,Freer:1994zz}; thus, it is especially interesting to compare our results in a medium with conventional approaches. Notice that the effects in a medium might be present in ref. \cite{Raduta:2011yz} and this might explain the discrepancies from conventional approaches \cite{Raduta:2011yz,Rana:2013hka,Manfredi:2012zz,DellAquila:2017ppe,Smith:2017jub,Kirsebom:view2017,Kirsebom:2012zza,Itoh:2014mwa,Freer:1994zz}. We define the decay probability as Eq.(\ref{SumE}):
\begin{small}
\begin{equation}
\prod(E^*,\delta E)= \frac{\sum_{ij}(Y_R(DDE\,or\,LD,E_{ij})-Y_M(DDE\,or\,LD,E_{ij}))}{\sum_{ij}(Y_R(SD,E_{ij})-Y_M(SD,E_{ij}))},
\label{SumE}
\end{equation}
\end{small}
where the sum is extended over all relative kinetic energies corresponding to a $^{12}$C level with excitation energy E$^*$ from Eq.(\ref{Estar}) and variance $\delta$E, which we will vary to the smallest values allowed by the statistics. The Y$_{R}$(SD, E$_{ij}$) and Y$_{M}$(SD, E$_{ij}$) in the denominator are obtained by fixing the smallest relative kinetic energy to the $^{8}$Be$_{g.s.}$ $\pm$ $\delta$E/3 for the real (R) and mixing (M) events, respectively. The yields of DDE or LD are obtained by opportunely choosing the relative kinetic energies in the numerator. For example, the DDE case is obtained by choosing three equal relative kinetic energies (within $\delta$E/3 for each one). For a fixed excitation energy, we can estimate Eq.(\ref{SumE}) from the data by changing $\delta$E in order to derive the limiting value of the ratio compatible with the experimental sensitivity.

\begin{figure}[!h]
\vspace{-5pt}
\setlength{\abovecaptionskip}{0.cm}
\setlength{\belowcaptionskip}{-0.cm}
\centering
\includegraphics[scale=0.4]{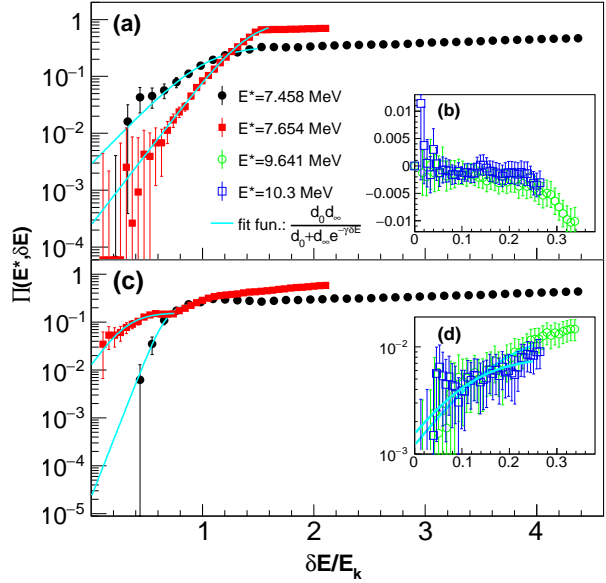}
\caption{\footnotesize (color online) The probability ratios of the DDE (top) and LD (bottom) as a function of the error (normalized to the kinetic contribution to E$^*$ from Eq.(\ref{Estar})).}
\label{Ratios}
\vspace{-5pt}
\end{figure}

In Fig. \ref{Ratios}, we plot the ratios for the DDE (top panel) and the LD (bottom panel) as a function of $\delta$E (normalized to the kinetic contribution to E$^*$ from Eq.(\ref{Estar})). The ES (E$^*$=7.458 MeV) and the HS (E$^*$=7.654 MeV) cases are given, respectively, by the solid (black) circles and the solid (red) squares. We have divided the ES cases by the HS SD, thus these ratios give the probabilities of decay of the ES with respect to the HS. The 9.641 MeV (green open circles) and the 10.3 MeV (blue open squares) are given in the insets. The ratios can be reproduced by the function \cite{chaos} given by Eq.(\ref{ratioeq}):
\begin{small}
\begin{equation}
\prod(E^*,\delta E)= \frac{d_0d_{\infty}}{d_0+d_{\infty}e^{-\gamma\delta E}}.
\label{ratioeq}
\end{equation}
\end{small}
\quad~The parameter $d_0$ ($\delta$E$\rightarrow$0) gives the smallest possible physical value of the ratios or the experimental error, while the largest value $d_{\infty}$ ($\delta$E$\rightarrow$$\infty$) is connected to the available phase space \cite{chaos}. The fit values of $d_0$ are reported in Table ~\ref{tab2} and are compared to other literature data. Since ref. \cite{Raduta:2011yz} might contain effects from within a medium as in our case, we argue that the difference is due to not properly subtracting the mixing events when calculating the ratios, as in Eq.(\ref{SumE}). Another possibility is the contribution of the ES due to the experimental sensitivity. We can see that the LD contribution of the ES is compatible with zero, which is consistent with the definition of the ES. For the larger excitation energies considered here, the ratios are negative for the DDE case (see Fig. \ref{Ratios}(b)) and compatible with zero for the LD case (see Fig. \ref{Ratios}(d)).

\begin{figure}[h]
\vspace{-5pt}
\setlength{\abovecaptionskip}{0.cm}
\setlength{\belowcaptionskip}{-0.cm}
\centering
\includegraphics[scale=0.55]{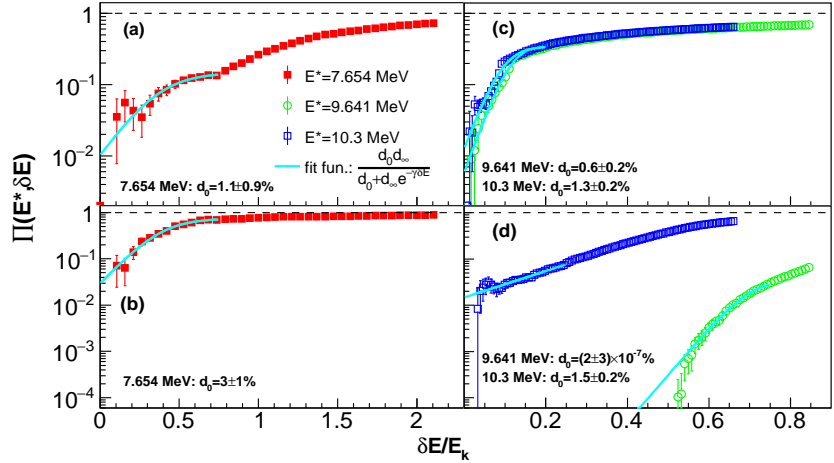}
\caption{\footnotesize (color online) The probability ratios of the HS (a) when the $^{8}$Be$_{g.s.}$ energy is consistent with two $\alpha$ relative kinetic energies, or (b) the second largest energy is twice of the $^{8}$Be$_{g.s.}$ energy. The ratios are plotted for 9.641 MeV (green open circles) and 10.3 MeV (blue open squares), (c) when the E$^{Min}_{ij}$=0.092 MeV, E$^{Mid}_{ij}$=E$^{Lar}_{ij}$, or (d) the E$^{Lar}_{ij}$=3.03 MeV, E$^{Min}_{ij}$+E$^{Mid}_{ij}$=$\frac{3}{2}\times$(E* + Q) - E$^{Lar}_{ij}$.}
\label{SD}
\vspace{-5pt}
\end{figure}

\begin{table}[!b]
\vspace{-5pt}
\setlength{\abovecaptionskip}{0.cm}
\setlength{\belowcaptionskip}{-0.cm}
   \centering
   \scriptsize
   \tabcolsep=1.5pt
   \caption{\footnotesize Fit values of the parameter $d_0$ compared to
   refs. \cite{Raduta:2011yz,Rana:2013hka,Manfredi:2012zz,Kirsebom:2012zza,Itoh:2014mwa,Ishikawa:2014mza,Zheng:2018plb,Smith:2017jub}}
   \vspace{2pt}
   \begin{tabular}{|c|c|c|c|c|c|c|c|c|}

   \hline 
   &ES:DDE(\%)& ES:LD(\%)& HS:DDE(\%)& HS:LD(\%)& 9.641:DDE(\%)& 9.641:LD(\%)& 10.3:DDE(\%)& 10.3:LD(\%)\\
   \hline
   Present&  0.3$\pm$0.1 &  0.002$\pm$0.004 & 0.025$\pm$0.005 & 1$\pm$1 & 0 & 0.1$\pm$0.1 & 0 & 0.2$\pm$0.2 \\
   \hline
   ref.\cite{Raduta:2011yz}&     &        & 7.5$\pm$4.0  & 9.5$\pm$4.0 & 0 & & &  \\
   \hline
   ref.\cite{Rana:2013hka} &     &        & 0.3$\pm$0.1  &    0.1      &  & & &   \\
   \hline
   ref.\cite{Manfredi:2012zz}&   &        &    0.45     &              &  & & &   \\
   \hline
   ref.\cite{Smith:2017jub}  &   &        &    0.036     & 0.024       &  & & &   \\
   \hline
   ref.\cite{Kirsebom:2012zza}&  &        &    0.09      &    0.09     &  & & &   \\
   \hline
   ref.\cite{Itoh:2014mwa} &     &        &    0.08      &             &  & & &   \\
   \hline
   ref.\cite{Ishikawa:2014mza}&  &        &    0.005     &    0.03     &  & & &   \\
   \hline
   ref.\cite{Zheng:2018plb}&     &        &    0.0036     &            &  & & &   \\
   \hline 
   \end{tabular}
   \label{tab2}
   \vspace{-5pt}
   \end{table}

The decay modes results are consistent with the quantity F in Eq.(\ref{yield}), which confirms that the $^{12}$C decay is intimately connected to the formation of $^{8}$Be, as illustrated by Figs. \ref{yield_2} and \ref{yield_3}. Not only the excited levels of $^{8}$Be determine the SD modes of $^{12}$C but also multiple integers of the $^{8}$Be$_{g.s.}$ energy. In the left panels of Fig. \ref{SD}, we have plotted the ratios for the HS when the $^{8}$Be$_{g.s.}$ energy is consistent with E$^{Min}_{ij}$ and E$^{Mid}_{ij}$ (Fig. \ref{SD}a) or the E$^{Mid}_{ij}$ is twice of the $^{8}$Be$_{g.s.}$ (Fig. \ref{SD}b). In the right panels of Fig. \ref{SD}, the ratios are plotted for 9.641 MeV and 10.3 MeV, when the E$^{Min}_{ij}$=0.092 MeV, E$^{Mid}_{ij}$=E$^{Lar}_{ij}$
(Fig. \ref{SD}c), or the E$^{Lar}_{ij }$=3.03 MeV, E$^{Min}_{ij}$+E$^{Mid}_{ij}$=$\frac{3}{2}\times$(E* + Q) - E$^{Lar}_{ij}$ (Fig. \ref{SD}d). There are decay modes with those
conditions, in particular, when the relative kinetic energies are approximately equal to the ratio 1:2:3, a (3$\pm1$)$\%$ contribution to
SD results for the HS, see Fig. \ref{SD}b.


In conclusion, in this paper, we have discussed energy levels of $^{8}$Be and $^{12}$C in hot nuclear matter. We found that the DDE and LD decay modes are strongly depleted. Thus the decay probability is mainly determined by the $^{8}$Be formation probability in $^{12}$C. Depending on the excitation energy of $^{12}$C, $^{8}$Be might be formed, not only in the ground state, but also in excited states as well. We confirm the finding of ref. \cite{Tumino:2015jaa} that some decay modes are dominated by $^{8}$Be levels hit more than once. A special case is the ES when the relative energies of 3$\alpha$s are consistent with the $^{8}$Be$_{g.s.}$, a signature of a strongly resonating Boson gas or an Efimov state, consistent with observations in atomic systems refs. \cite{Zaccanti:2009NP}. Some DDE and LD decay modes might be observed at very large excitation energies and these will be discussed further in a following work together with the question of BEC.

\section*{ACKNOWLEDGMENTS}
This work was supported by the National Natural Science Foundation of China (No. 1176014, 11605097, 11421505), the US Department of Energy under Grant No. DE-FG02-93ER40773, NNSA DE-NA0003841 (CENTAUR) and the Robert A. Welch Foundation under Grant No. A-1266. This work was also supported by the
Chinese Academy of Sciences (CAS) President's International Fellowship Initiative (No. 2015VWA070), Strategic Priority Research Program of the Chinese Academy of Sciences (No. XDB16), Program for Young Talents of Science and Technology in Universities of Inner Mongolia Autonomous Region (NJYT-18-B21) and Doctoral Scientific Research Foundation of Inner Mongolia University for Nationalities (No. BS365, BS400).
AB thanks the Chinese Academy of Science, Sinap and Inner Mongolia University of Nationalities for the warm hospitality
and support during his stay in China while this work was completed.








\end{document}